\documentclass{article}

    \PassOptionsToPackage{numbers, compress}{natbib}
   \usepackage[final]{neurips_2021}

\usepackage[utf8]{inputenc} % allow utf-8 input
\usepackage[T1]{fontenc}    % use 8-bit T1 fonts
\usepackage{hyperref}       % hyperlinks
\usepackage{url}            % simple URL typesetting
\usepackage{booktabs}       % professional-quality tables
\usepackage{amsfonts}       % blackboard math symbols
\usepackage{nicefrac}       % compact symbols for 1/2, etc.
\usepackage{microtype}      % microtypography
\usepackage{xcolor}         % colors
\usepackage{url}

\title{How Can Creativity Occur in Multi-Agent Systems?}

\author{%
  Ted Fujimoto \\
  Computing and Analytics Division\\
  Pacific Northwest National Laboratory\\
  Seattle, WA 98109 \\
  \texttt{ted.fujimoto@pnnl.gov} \\
}

\begin{document}

\maketitle

\begin{abstract}
Complex systems show how surprising and beautiful phenomena can emerge from structures or agents following simple rules. With the recent success of deep reinforcement learning (RL), a natural path forward would be to use the capabilities of multiple deep RL agents to produce emergent behavior of greater benefit and sophistication. In general, this has proved to be an unreliable strategy without significant computation due to the difficulties inherent in multi-agent RL training. In this paper, we propose some criteria for creativity in multi-agent RL. We hope this proposal will give artists applying multi-agent RL a starting point, and provide a catalyst for further investigation guided by philosophical discussion.
\end{abstract}

\section{Introduction}

The title of this paper takes inspiration from Allen Newell's scientific question: ``How Can the Mind Occur in the Physical Universe?" \citep{newell1993desires}. \citet{anderson2009can}, a book motivated by the same question, proposes ``cognitive architecture" as a framework that describes the structure and function of cognition. We will take a similar approach to creative, emergent behavior in multi-agent RL. In particular, we propose a non-exhaustive list of guidelines for creative behavior in multi-agent systems. These guidelines are an attempt to understand the structure and mechanisms of creative multi-agent RL.

The goal of this paper is to propose some steps forward for artists to utilize multi-agent RL for creative purposes. The balance between full and human-assisted automation is a difficult task, even for RL researchers, and finding that balance is still an open research question. There is also the reality that RL does not give humans an intuitive sense of direct control. An analogy would be a director telling the actors what to do, but allowing them the freedom to do it their own way. Multi-agent RL goes a step further by only specifying the environment rewards and letting the agents learn how to maximize those rewards on their own.

In the next section, we provide some RL concepts that could facilitate the emergence of creative group behavior. The main idea is to explicate multi-agent creativity as exploration by a diverse group of agents that learn to cooperative guided by limited human imitation. We will use Agence by \citet{camarenaagence} as a helpful guidepost for our criteria. The project's impressive contributions to the topic will give some background to our criteria. We will also use the reasons why they had to scale back on RL \citep{camarena2020Video} as a clear problem for our criteria to solve.

\section{Criteria for Creative Multi-Agent RL}

\emph{Individual exploration}: In single-agent RL, the problem of exploration is important for finding the optimal policy. This has been typically done by maximizing the entropy of the policy in some way. \citet{eysenbach2018diversity} showed that maximizing an information-theoretic objective with a maximizing entropy policy leads to an unsupervised emergence of diverse skills. \citet{zhang2021exploration} use a reward-free RL algorithm to first explore by maximizing R\`enyi entropy, and then use the reward function during the planning stage (e.g. batch RL) to learn the optimal policy. In a way, a policy that maximizes the cumulative policy entropy at each state is searching for ways to avoid the terminal state (like the agent's death). The reason is the following: if we assign the terminal states to have zero entropy, then states that are close (in terms of the number of actions needed to reach it) to many terminal states will have less state-value. This implies states with high cumulative policy entropy are ``far away" from terminal states. This can be interpreted as more intelligent exploration of the environment. A similar argument can be made for continuous action spaces by maximizing the variance of a normal distribution.
 
\emph{Group diversity}: If we want more interesting behaviors, we want the agents to exhibit behaviors dissimilar from one another. Although one can accomplish more when working in a team, there is a danger to creativity when all agents think in the same way \citep{nemethbetter}. This is commonly called \emph{groupthink}. For example, we cannot observe the combination of individual complex behaviors in Agence if the policies are similar and predictable. Static, unchanging policies also means it is less likely for some agents to exhibit behaviors that benefit the entire group. A principled way would be to optimize the diversity of all agents simultaneously. One way of doing this would be to extend effective diversity of population based RL \citep{parker2020effective} to multi-agent settings.

\emph{Efficiently learning human imitation}: Needless to say, the artist requires some level of control. The goal of imitation learning is for the agent to learn a policy that matches an expect given demonstrations of the expert's behavior. Due to the highly complex environments, real-world RL applications (e.g. Agence \citep{camarena2020Video}), tend to run into stagnant training that forces the designers to scale back the self-taught RL features in favor of some hard-coded behaviors. Past earlier attempts to solve this problem, like inverse RL \citep{ng2000algorithms} and behavioral cloning \citep{pomerleau1989alvinn}, have had difficulties reproducing expert-level behavior in real-world settings. More recent work has made progress in this area. For example, \citet{jaegle2021imitation} propose imitation learning solely from observations that achieves comparable performance to experts. Other than performance, another practical benefit is the designer not needing to account for the agent's low-level actions that might be intractable for humans.

\emph{Cooperative AI as an accelerator for cooperation and a test for defection}: The are multi-agent RL methods that are intended to make learning cooperation between agents more efficient. The hope is that these methods will allow for better coordination and social welfare than training multiple agents using single-agent RL methods \citep{dafoe2020open}. For example, Agence requires the group to work together to coordinate control of their planet to avoid falling to their doom. These methods can also be a test for avoiding scenarios that have a high chance of leading to agent defection. The human designers can see if cooperative AI methods consistently fail and lead to the destruction of the group in certain cases. This may involve observing the \emph{social behavior metric} from \citet{leibo2017multi} to measure when an agent's policy is defecting.

\section{An Example of Creative Multi-Agent RL}
In an environment like Agence, you try to avoid monotonous cooperation over a long period of time, but you also want to avoid agents constantly defecting and killing each other. The ideal scenario is watching the diverse group of agents try to cooperate amid the changes in the environment. We could train the group as follows: (1) At the initial stage, use entropy maximization and effective diversity so that all agents explore and learn how to survive in different ways without knowledge of the reward, (2) use a cooperative AI algorithm (e.g.  Cooperative Approximate Policy Iteration \citep{sokota2021solving}) that facilitates group cumulative reward maximization, and (3) use imitation learning to help the agents learn human-guided behaviors.

\section{Conclusion and Future Directions}
We have provided some criteria for creative multi-agent RL and some guidelines on how to apply it. One direction forward is to verify if this strategy works. Another direction would be to discuss what to add or remove from the list of criteria. For artists, we hope that we have illuminated the balance between automating the low-level, obscure details of RL creativity and the intuitive control to guide the agents' behavior.

\medskip

\small

\bibliographystyle{plainnat}
\bibliography{neurips}

\end{document}